\documentclass[twocolumn]{aastex631}

\usepackage{xcolor}
\usepackage{amsmath}

\shorttitle{Dawn of the Milky Way disk. I.}
\shortauthors{Feltzing et al.}

\begin{document}

\title{Dawn of the Milky Way disk: Determination of when a rotationally supported disk appears and dating the spin-up of the disk}

\author[0000-0002-7539-1638]{Sofia Feltzing}
\affiliation{Lund Observatory, Department of Earth and Environmental Sciences , S\"olvegatan 12, SE-223\,62 Lund, Sweden}
\affiliation{Lund Observatory, Department of Geology, S\"olvegatan 12, SE-223\,62 Lund, Sweden}
\email{sofia.feltzing@mgeo.lu.se}

\author[0000-0002-3101-5921]{Diane Feuillet}
\affiliation{Observational Astrophysics, Department of Physics and Astronomy, Uppsala University, Box 516, SE-751 20 Uppsala, Sweden}

\author[0000-0003-3978-1409]{Thomas Bensby}
\affiliation{Lund Observatory, Division of Astrophysics, Department of Physics, Lund University, Box 118, SE-221\,00 Lund, Sweden}



\begin{abstract}
Spiral galaxies, like the Milky Way, transform at some point in time into a rotationally supported system. Using an extant data-set consisting of 319\,835 sub-giants from LAMOST with precise ages from the literature, we determine, for the first time the age when the Milky Way disk spins up, i.e. when the mean circular velocity changes from halo-like to disk-like. We find in concordance previous studies that the spin-up takes place for  $-1.25<$[Fe/H]$<-0.9$ and we can date this transition to a mean age of  12.1$\pm 2.8$\,Gyr (median age 12.4\,Gyr). 
We further study when the disk became rotationally supported, i.e. when  the ordered, disky motion dominates over the random motions. We find that this happens for $-1.25<$[Fe/H]$<-1$. The transition is very rapid in age. This gives support to that the spin-up seen in this and other works genuinely traces the motion to a rotationally supported disk, which has not previously been shown. 
These transitions are traced by the high-$\alpha$ stars
while the low-$\alpha$ stars do not spin-up but start directly at approximately the circular velocity seen for the Sun today. The low-$\alpha$ disk is rotationally supported with no transition period  in [Fe/H] or in age.

\end{abstract}



\section{Introduction} \label{sect:intro}

When and how galaxies form is a fundamental question to answer for understanding the evolution of the Universe. Recent years have, thanks to James Webb Space Telescope (JWST) and Atacama Large Millimeter Array (ALMA), seen a surge of high redshift observations indicating that rotationally supported disky galaxies are in place already at redshift 6-7, grand design spiral galaxies around $z \sim 4$, and very massive galaxies at $z\sim 5.2$\citep[e.g.,][]{2023MNRAS.521.1045R,2024MNRAS.535.2068R,2025A&A...703A..96J,2025A&A...696A.156X}. To quote \citet{2023ApJ...955...94F} “While earlier HST-based studies found that the dominating majority of galaxies at z $> 2$ are peculiar, recent JWST-based studies find a high number of regular disk galaxies at high redshift". This poses a significant shift in our understanding of the timeline of the build up of galaxies in the Universe. 
Being a  disk in a galaxy (stellar or gaseous) means being rotationally supported, which means that the rotational motion is dominating over the random motions.  
Finding out when galaxies started to host rotationally supported stellar disks thus becomes interesting to study also in the local Universe. Here, the Milky Way provides an excellent, but challenging, testbed.

How can we trace the moment when the Milky Way started to host a rotationally supported stellar disk (i.e. $V_{\phi}/\sigma_{vel.}>1$)? Observationally we have access to kinematics, metallicities and ages of the stars in the Milky Way today. We can thus study velocities and velocity dispersions as a function of stellar age but keeping in mind that stellar orbits can be peturbed these observations provide a minium age at which the Milky Way was rotationally supported.

 Instead, the period known as the spin-up, i.e. the period when the kinematic signature transitions from one dominated by a spherical halo to one dominated by a disk, has gained attention. Several studies have looked at which metalliity the overall kinematics of the Milky Way changes from $<V_{\phi}> \sim 0$ to $<V_{\phi}> \sim 200 - 220$\,km\,s$^{-1}$. This transition has been studied in observational data   \citep[including][]{2022MNRAS.514..689B,2024A&A...688A.167N,2022arXiv220402989C,2024arXiv241112165V,2025arXiv250909576O} and in simulations \citep[including][] {2024MNRAS.527.6926M,2024MNRAS.527.7070D,2025arXiv250611840M,2025ApJ...990....7S,2025arXiv250909576O}. Simulations tend to find that Milky Way-like galaxies spin-up at metallicities higher than what is found observationally for the Milky Way.

The period  when the Milky Way disk  become rotationally supported is a more robust measure than the spin-up, which is somewhat poorly defined, and can be the result of a coincidence of stellar populations, while observation of a rotationally supported disk is a genuine result and cannot be a coincidence. 

The observational focus on  [Fe/H] as the reference rather than stellar age is partly necessitated by the fact that [Fe/H] is  readily  measurable for large numbers of stars while ages for massive data-sets have only recently become available. Stellar ages are more straightforward to interpret than [Fe/H] and within a given observational data-set or a simulation it is safe to assume that the ordering of the stars along the age-axis is robust \citep{2024A&A...692A.243C}.
In our work we wish to exploit the new possibilities provided by major surveys including turn-off and sub-giant stars in large numbers. 
 
 The straightforwardness of interpretation of age in simulations is not echoed by the metallicities, which
 in contrast, is highly dependent on the modelling of the chemical evolution, including in- and out-flows.

This paper is organized as follows: Section\,\ref{sect:data} describes the data we are using and how we select the sample. In Sects.\,\ref{sect:rotsupp} and \ref{sect:dating} we determine when the Milky Way becomes rotationally supported and when it spins-up. Finally, Sect.\,\ref{sect:sum}  summarizes and discusses our results.

\section{Data}\label{sect:data}

We use the data from \citet{2025NatAs...9..101X}, which provides an update of \citet{2022Natur.603..599X}. The catalogue consists of stellar parameters ($T_{\rm eff}$, $\log g$, [Fe/H], [$\alpha$/Fe], and ages) and kinematic data for   320\,028 sub-giant stars. The reader is referred to \citet{2025NatAs...9..101X} for details on how each parameter was derived and in Appendix\,\ref{app:data} we give further details and links to catalogues for this data set. In that section and in Appendix\,\ref{app:cuts}  we outline how we investigate the quality of the data and the reasons for various quality cuts. The latter are as follows:

\begin{itemize} 
	\item {\tt RUWE} (Renormalized Unit Weight Error),  which is a measure of the goodness of the $\chi^2$ of the astrometric fit for {\it Gaia}.\footnote{L. Lindegren (2018) Renormalizing the astrometric chi-square in Gaia DR2. Lund Observatory. Note: Gaia Data Processing and Analysis Consortium (DPAC) technical note GAIA-C3-TN-LU-LL-124, \url{http://www.cosmos.esa.int/web/gaia/public-dpac-documents} External Links: Link Cited by: 20.1.1.} We remove stars with {\tt RUWE} $>$ 1.4.
	\item We check the error in parallax and a cut of 20\% relative error appears reasonable.
 	\item We apply a hard cut at $\log g \geq3.2$ as this removes the majority young metal-poor stars.
	\item We apply a hard cut at $T_{\rm eff}  \leq 6500$\,K to remove stars with poor age determinations.
\end{itemize}

We kinematically remove  stars belonging to {\it Gaia}-Enceladus in order to avoid confusion between the existing Milky Way stars  and a major accreted stellar population with distinct kinematic properties, see discussion in Appendix\,\ref{app:gse}.

Finally, we decided to only keep stars with $|z| < 1.5$\,kpc. By cutting in $z$ we ensure that the results are relevant for the Milky Way disk population and  remove contamination from the halo which would tend to put the spin up at higher [Fe/H], as the halo has low metallicities and a non-rotating kinematic signature. See also figures in Appendix\,\ref{app:cuts} and \ref{app:zcut}.

\section{A rotationally supported disk}
\label{sect:rotsupp}

\begin{figure*}
\includegraphics[width=18cm]{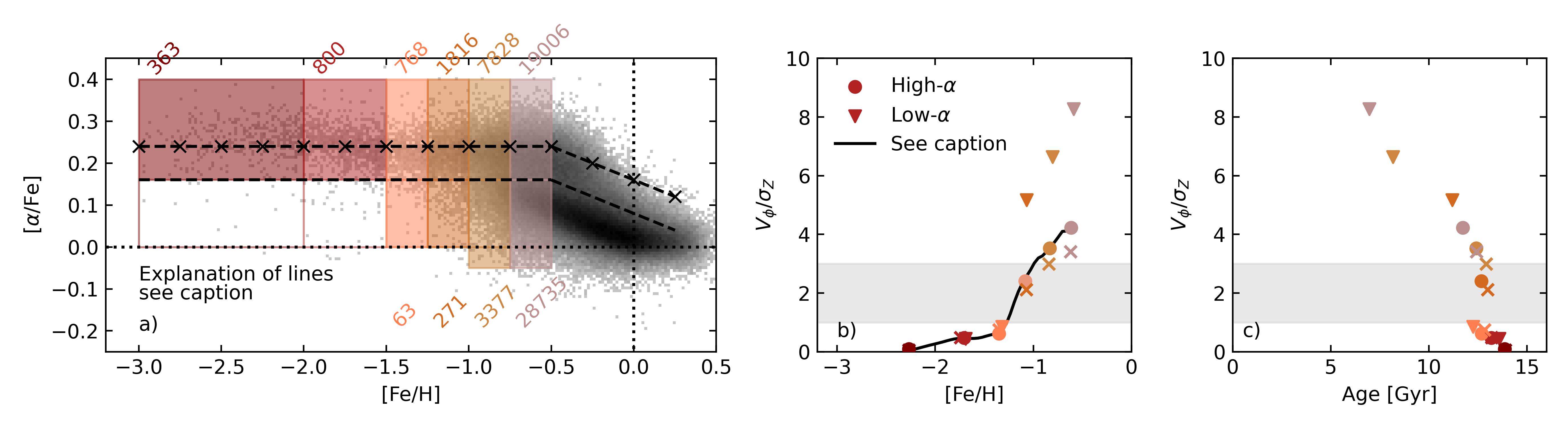}
\caption{{\bf a)} [$\alpha$/Fe] as a function of [Fe/H]  for the full sample, with {\it Gaia}-Enceladus stars removed (see Sect.\,\ref{sect:data}). The underlying data are shown as a 2D histogram with power-law relationship$^{\ast}$. The coloured boxes define the different subsamples analysed in panels b and c. The number of stars in each bin is indicated in. The dashed line shows Eq.\,(\ref{eq:alpha_choice}) which splits the data into high- and low-$\alpha$ stars and the dashed line with crosses the same line but raised by 0.08\,dex. {\bf  b)} The median $V_{\phi}/\sigma_{Z}$ as a function of the median [Fe/H] for stars in each of the bins defined in panel a. High- and low-$\alpha$ stars split by Eq.\,(\ref{eq:alpha_choice}) as indicated and high-$\alpha$ stars resulting from the raised cut are shown by $\times$. The full line shows a smoothed fit to the underlying data. {\bf c)} The median $V_{\phi}/\sigma_{Z}$ as a function of the median age for stars in each of the bins defined in panel a.  High- and low-$\alpha$ stars split by Eq.\,(\ref{eq:alpha_choice}) as indicated and high-$\alpha$ stars resulting from the raised cut are shown by $\times$. The full line shows a smoothed fit to the underlying data. The grey shaded areas in b) and c) show the region $1< V_{\phi}/\sigma_{\rm Z} <3$, see discussion in Sect.\,\ref{sect:rotsupp}.\newline
$^{\ast}$\footnotesize{This means that the  colors are remaped onto a power-law relationship, $y=x^{\gamma}$ where ${\gamma}$ is the power and if it is equal to 1 it simply gives the default normalization in Matplotlib.}}
\label{fig:rotsup}
\end{figure*}

Stars share their motions with the gas cloud from which they form. When the circular velocity becomes larger than the random motions in the gas disk, including flows due to stellar feedback, stars can form \citep[][]{2025ApJ...990....7S}.  The gas can later obtain a different motion and so can the stars. For example stars can get heated orbits via interactions with a bar or via a merger \citep[][and references therein]{2025arXiv250909576O}. We measure the motions of the stars today. This means that we measure their motion at formation plus any interactions that change their orbits before observation. A  stellar disk is said to be rotationally supported if the rotational velocity ($V_{\phi}$) of the stars is larger than their random motions ($\sigma_{\rm vel.}$), i.e. $V_{\phi}$/$\sigma_{\rm vel.} > 1$. If $V_{\phi}$/$\sigma_{\rm vel.} > 1$ today, we know that the system was rotationally supported at the time of  formation. However, a system might have been initially rotationally supported but subsequently heated such that it no longer appears rotationally supported, i.e. $V_{\phi}$/$\sigma_{\rm vel.}<1$. 

In most instances a threshold of 1 is used \citep[see for example][]{2025MNRAS.543.3249D} to signify that a stellar population or gas disk is rotationally supported, but practices differ and values of $V_{\phi}$/$\sigma_{\rm vel.}=$1.5 or 3 can be found in the literature. 

Next we need to decide which velocity dispersion to use. The literature is not fully consistent on this point. In some works $\sigma_{\rm Tot}$ is used, while in others $\sigma_Z$ is used. It is well established that in the Milky Way $\sigma_{\rm R} >  \sigma_{\phi} > \sigma_{\rm z}$ \citep{2009A&A...501..941H} and that the velocity dispersions increases with stellar age \citep{2025A&A...700A..89K}. The three velocity dispersions are influenced by different physical processes; $\sigma_{\rm R}$ and $\sigma_{\phi} $ are essentially sensitive to processes in the plane such as interactions with spiral arms and bars, while $\sigma_{\rm z}$ reflects only heating in the vertical direction which is provided e.g. by giant molecular clouds but also more major events such as the Splash \citep{2025A&A...700A..89K}.  We choose to use $\sigma_{\rm Z}$ since it traces the vertical heating. However, in order to reflect that indeed this is only one of the velocity dispersions we use $1< V_{\phi}$/$\sigma_{\rm z}<3$ as our outer boundaries for the disk to transition to rotationally supported. A value of $V_{\phi}$/$\sigma_{\rm vel.}=$1 would be appropriate for a disk which has no heating radially or in azimuth while the value of $V_{\phi}$/$\sigma_{\rm vel.}=$3 could indicate all three dispersions are the same strength, which is often assumed in work on other galaxies. We show this range as a grey band in Figs.\,\ref{fig:rotsup} b) and c).

We  divide the sample into high- and low-$\alpha$ samples. The cut between the two samples is given by Eq.\,(\ref{eq:alpha_choice}), which is adapted from \citet{2022Natur.603..599X}: 

\begin{equation}
  \begin{cases}
    [\alpha/\rm Fe] > 0.16, & {\rm if \,[Fe/H]} < -0.5 \\
    [\alpha/\rm Fe]  > -0.16 \cdot{\rm [Fe/H]} + 0.08, & {\rm if \,[Fe/H]} > -0.5, \\
    \end{cases} 
\label{eq:alpha_choice}
\end{equation}

The division is less clear as we go to higher metallicities, but as we are mainly concerned with stars below --0.5 dex and hence we have  a straight line cutting the sample in two with different [$\alpha$/Fe], see Fig.\,\ref{fig:rotsup}a) where the line is shown. We further divide the sample into bins of [Fe/H]  and [$\alpha$/Fe], see Fig.\,\ref{fig:rotsup}a). For each bin, we calculate $V_{\phi}$/$\sigma_{\rm Z}$. These are shown as a function of [Fe/H] in Fig.\,\ref{fig:rotsup}b). 

We find that the high-$\alpha$ stars reaches $V_{\phi}$/$\sigma_{\rm z}=1$ at   [Fe/H] =  --1.25 and the trend then rises, reaching  $V_{\phi}$/$\sigma_{\rm vel.}=$3 at [Fe/H] = --1. We thus determine that the high-$\alpha$ stars move from a not rotationally supported to an undoubtedly rotationally supported configuration over about 0.25 dex in [Fe/H], while the low-$\alpha$ stars  start out rotationally supported from the lowest metallicities where they are abundant, above [Fe/H]$=-1.25$\,dex. 

Next, we look at the ages of this transition. For each of the bins defined in Fig.\,\ref{fig:rotsup}a) we calculate their median age and plot $V_{\phi}$/$\sigma_{\rm Z}$ as a function of that age in Fig.\,\ref{fig:rotsup}c). Here the transition is even more rapid. Specifically, the high-$\alpha$ stars move from a $V_{\phi}$/$\sigma_{\rm Z}$ 1 to  $V_{\phi}$/$\sigma_{\rm Z}$ of 3 almost instantly.

\section{Dating the spin-up}
\label{sect:dating}

\begin{figure*}
\includegraphics[width=15cm]{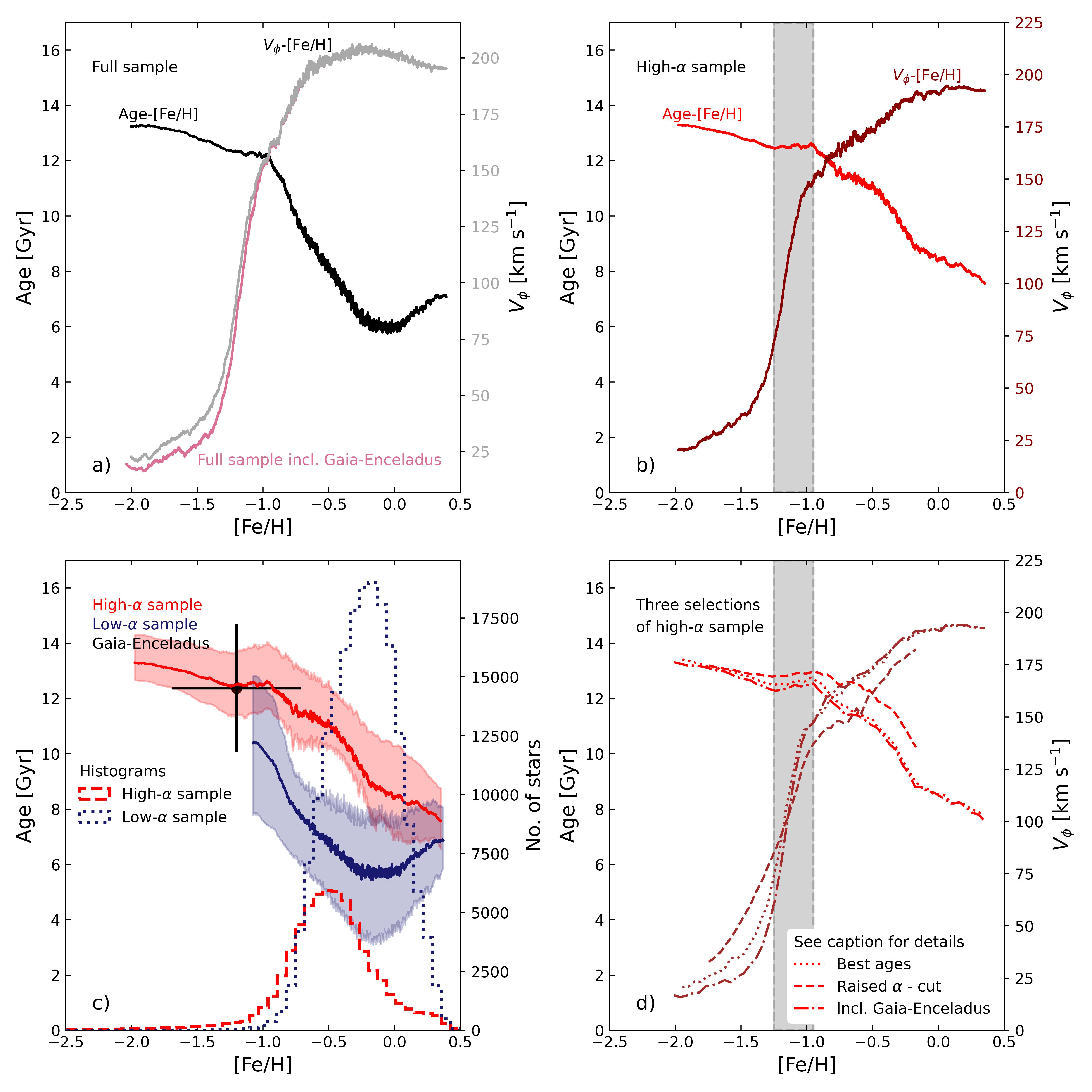}
\caption{Determination of the age of the spin-up. The trends are calculated by rolling over a bin including  1000 data-points along the [Fe/H]-axis. {\bf a)} Full sample. Age -- [Fe/H] relation shown in black and $V_{\phi}$ -- [Fe/H] relation in grey. {\bf b)} High-$\alpha$ sample. $Age$ -- [Fe/H] relation shown in red and $V_{\phi}$ -- [Fe/H] relation in brown. The grey shaded area shows the identified [Fe/H] range of the spin-up. {\bf c)} Comparison of the $Age$ -- [Fe/H] relations for the high- and low-$\alpha$ samples in red and blue, respectively. The red and blue shaded areas show the inter-quartile ranges for each sample. The median position of the kinematically selected {\it Gaia}-Enceladus stars in [Fe/H] and age are shown as a black filled circle with error bars indicating the one sigma standard deviation. The metallicity distributions for the  high- and low-$\alpha$ samples are shown as histograms in red and blue, respectively. {\bf d)} Same as b, but applying more restricted selection criteria. ``Best age" means an error in $Age$ of less than 10\%, ``Raised $\alpha$ - cut" indicates a sample where the cut for the high-$\alpha$ sample has been raised by 0.08\,dex. This means that the sample stops just below solar metallicity. In this panel, all three curves are calculated rolling over a bin of 1000 data points,  but here the curves have also been smoothed to ease the comparison. Line-styles as indicated in the legend.  }
\label{fig:spinup}
\end{figure*}

The term spin-up ordinarily applies to the transition phase when the stars progress from random motion to ordered, disky motion as measured by $<V_{\phi}>$ \citep{2022MNRAS.514..689B,2024MNRAS.527.6926M,2024ApJ...972..112C,2024arXiv241112165V,2025ApJ...990....7S}. Our analysis for determining the age of the spin-up of the Milky Way disk is shown in Fig.\,\ref{fig:spinup}. To obtain the age of the spin up we compare the $V_{\phi}$-[Fe/H] and $Age$-[Fe/H] trends. These trends were obtained by  ordering  the data in [Fe/H] and calculate  $V_{\phi}$ and Age using a rolling window of 1000 data points. We have checked that the exact choice of bin-size does not impact the results.

Figure\,\ref{fig:spinup}a) shows the resulting plot for the full sample: the Age -- [Fe/H] trend is first slowly declining, then stalls, and then decreases rapidly, reaching a minimum around [Fe/H]$\sim -0.2$ and then rising again. The $V_{\phi}$-[Fe/H] trend starts low, then rises rapidly to a peak around [Fe/H] $\sim -0.3$ and then gently decline towards the highest [Fe/H]. We note that the stalling of the Age -- [Fe/H] trend coincides with the rise of the  $V_{\phi}$-[Fe/H] trend. Based on this plot we define a window in [Fe/H] when this stalling happens,  $-1.25 <$[Fe/H]$-0.9$.

Figure\,\ref{fig:spinup}b) shows the resulting plot for the high-$\alpha$ sample only. The spin-up window defined uabove is indicated. We find that the $Age$-[Fe/H]  trend for the high-$\alpha$ sample stalls at the same [Fe/H] as the full sample and that the rise of the $V_{\phi}$-[Fe/H] of the high-$\alpha$ sample falls in the same window. The  trends for the low-$\alpha$ sample are not shown but the trend starts at [Fe/H] $\sim -1$ where the $V_{\phi}$ is already 200 -- 220 km\,s$^{-1}$ and continues at this level for all [Fe/H]. Hence, there is no spin-up period (in age or [Fe/H]) associated with the low-$\alpha$ sample.

Figure\,\ref{fig:spinup} c) show that the $Age$-[Fe/H]  trends for the two samples are well separated in the range $-1$ to $-0.25$\,dex. 

We test the robustness of our result by checking three samples with different selection criteria. Selection 1: The best ages are ages with a formal error of less than 10\%. Selection 2: The higher cut in [$\alpha$/Fe] is defined as simply shifting the cut in Eq.\,(\ref{eq:alpha_choice}) 0.08\,dex higher, see Fig.\,\ref{fig:rotsup}. Selection 3: Leaving the {\it Gaia}-Enceladus in the sample. 

We find that the stalling of the $Age$ - [Fe/H] trend for all three selections is fully consistent with the result using the full high-$\alpha$ sample. That the three selections give slightly different ages for the plateau, but are consistent with each other and with that of the full sample within the quartiles shown in panel c). The $V_{\phi}$-[Fe/H] trend differ somewhat; with a steeper rise for the best ages and a slower ascent for the higher $\alpha$-cut \citep[compare][who also find that the exact cut in $\alpha$-abundance matters for the exact shape of the curve]{2025arXiv250909576O}. We confirm that our result is robust agains the method of selecting the stars. 

We find that the spin-up of the disk identified as the stars with $-1.25<$ [Fe/H] $< -0.9$  is $12.1\pm 2.9$\,Gyr (4381 stars) for the full dataset and $12.5\pm 2.5$\,Gyr (3657 stars) for the high-$\alpha$ sample (with median values of 12.5 and 12.7 Gyr, respectively). The low-$\alpha$ stars in that metallicity range are significantly younger with a mean age of 10.2 $\pm$3.5. All the stars kinematically identified as belonging to the {\it Gaia}-Enceladus  (see Sect.\,\ref{sect:data}) have a mean age of $12.3\pm 2.3$\,Gyr  (black cross in Fig. \ref{fig:spinup}c). This value includes all stars belonging to the {\it Gaia}-Enceladus regardless of metalllicity (3245 stars). We thus find that the stars in the  {\it Gaia}-Enceladus are slightly younger than the stars participating in the spin-up of the high-$\alpha$ disk. 

We further find that the spin-up iron abundances found here fully agree with the [Fe/H] range for which we find that the high-$\alpha$ disk becomes rotationally supported, i.e. rising above 1 and reaching 3 (see Sect.\,\ref{sect:rotsupp}) and that the low-$\alpha$ stars do not show any spin-up signature.
To conclude, we have, for the first time, put a time-stamp on the spin-up of the Milky Way as well as related that time-stamp to the age of the stars in the {\it Gaia}-Enceladus and the low-$\alpha$ stars.

\section{Summary and discussion}
\label{sect:sum}

Taking advantage of a large data-set of sub-giants with precise iron abundance, $\alpha$-elements and ages we have studied the period when the Milky Way stellar disk becomes rotationally supported and determined the time when it spun-up. We find that the high-$\alpha$ stars become a rotationally supported system over 0.25\,dex in [Fe/H]. This sharp rise in $V_{\phi}/\sigma_{\rm Z}$ over a short range in [Fe/H] takes place over an even shorter period in time as measured by the ages of the participating stars. This implies a rapid change in kinematic properties with rapid chemical evolution over a short period in time. We additionally find, that the spin-up takes place over a range of [Fe/H]  but the range varies depending on how the sample is selected and is therefore less robustly defined than the change in rotational support. We are, however, able to reliably determine the spin-up to happen 12.1 Gyr ago (full sample) or 12.5 Gyr ago (high-$\alpha$ sample). This is the first time this period as been estimated.

We find that the stars in the {\it Gaia}-Enceladus are slightly younger than the stars participating in the transformation of the Milky Way into a rotationally supported, spinning disk galaxy.  \citet{2025arXiv250909576O} reason that since the Milky Way has an old, $\alpha$-enhanced disk that is rotating, this provides constrains on the last major merger and estimate that the merger with the {\it Gaia}-Enceladus, the spin-up of the $\alpha$-enhanced disk and a star burst all co-inside at about 11 Gyr ago. Based on their simulation, \citet{2025ApJ...990....7S} conclude that if the Milky Way did not suffer any major impacts since the time of spin-up it would mean that a rotating stellar disk could have formed as early as $z \simeq 6-7$.  \citet{2024MNRAS.527.7070D} analyzed ARTEMIS simulations with and without a {\it Gaia}-Enceladus-like merger and found consistent results such that galaxies with merger products would have spun-up earlier. All fully consistent with our findings. 
 
 Interestingly, in our work, it is the $\alpha$-enhanced disk, aka the thick disk, that traces the quick period when the Milky Way becomes rotationally supported. The low-$\alpha$ disk instead starts out directly at the high rotational velocity that today characterizes the Milky Way disk. At least some simulations appear to support this picture, especially \citet{2025ApJ...990....7S} who find that the thick disk becomes rotationally supported quickly. Their work also highlights the differences between simulations. They analyze TNG and ART zoom-in simulations and find important differences; in particular the ART simulation has an early rapid phase of spin-up, which happens over about 0.7 Gyr, while the TNG simulation has a more gradual spin-up phase. The ART simulation spin-up is very similar to what we obtain.
 
\citet{2025arXiv250909576O} analyze Auriga simulations with respect to the spin-up as a function of [Fe/H] and age.
They conclude that the stellar kinematics at $z=0$ rarely can recover the real time at which a disk spins-up. The reason for this is the disruptions to stellar kinematics caused by massive radial mergers. This means that the time constraint possible to get from the data will be an upper limit, ie the disk could very well have formed earlier than indicated by the motions of stars today. It is therefore perhaps not too surprising that  \cite{2025arXiv250611840M} find the Milky Way to be a kinematic outlier when comparing to FIRE-2 simulations. However, examination of  our data and the results from \citet{2020AJ....160...43A} and \citet{2023MNRAS.521.1045R} suggest that the Milky Way should be considered "hotter" than the Milky Way data used by \cite{2025arXiv250611840M}. In fact, we note that the "old" Milky Way disk actually fits their simulations well. It would be interesting to further explore the kinematic properties of the Milky Way in the context of nearby galaxies, such as the SAMI sample, \citet{2024MNRAS.529.3446C}, which shows that the $\lambda_{R_e}$ parameter, which traces galactic rotation, for these galaxies correlates with age. 

Based on our findings, we propose to focus future studies in the metallcity range -2 to -0.7 dex. It appears urgent to attempt to disentangle what seems to be a genuine halo component that starts at high-$\alpha$ but potentially has a down-turn at metallicities somewhat higher that the {\it Gaia}-Enceladus. This is then mixing with a high-$\alpha$ disk that becomes rotationally supported over a short metallicity range and an even shorter age range. On top of this we have the {\it Gaia}-Enceladus and a low-$\alpha$ disk that is distinct in age and also distinct in kinematics in that it starts out as a fully fledged rotational disk without any spin-up or period when it becomes rotationally supported. Future large scale surveys such as the 4MIDABLE-HR with 4MOST \citep[][Bergemann, Bensby, et al. in prep]{2019Msngr.175...35B} will allow us to address this in depth and with a range of $s$- and $r$-process elements that would help us to better disentangle different stellar populations.


\begin{acknowledgments}

The authors thank the anonymous referee for helpful comments that improved the presentation of the investigation and results.
S.F. was supported by a project grant from the Knut and Alice Wallenberg Foundation (KAW 2020.0061 Galactic Time Machine). D.K.F. acknowledges funding from the Swedish Research Council grant 2022-03274. T.B. acknowledges funding from the Swedish Research Council grant 2024-04990.

\end{acknowledgments}

%






\appendix
\section{Data description including discussion on quality cuts and cuts in $z$}
\label{app:data}

All data used in this investigation are taken from \citet{2025NatAs...9..101X}, which provides an update of \citet{2022Natur.603..599X}\footnote{DOI:10.12149/101467 \citet{10.12149/101467}. We downloaded the data from \url{https://nadc.china-vo.org/res/r101467/}.}, which was based on the seventh data release (DR7) of the LAMOST Galactic survey \citep{2022Innov...300224Y}\footnote{\url{https://dr7.lamost.org}} combined with astrometric data from the {\it Gaia} satellite \citep[third data release,][]{2021A&A...649A...1G}. The catalogue consists of stellar parameters ($T_{\rm eff}$, $\log g$, [Fe/H], [$\alpha$/Fe], and ages) and kinematic data for   320\,028 sub-giant stars. We refer the reader to \citet{2025NatAs...9..101X} for details on how each parameter was derived, including the age determinations. 

We are interested in the precise ages in \citet{2025NatAs...9..101X}. They derived ages using Yonseii-Yale isochrones\footnote{\url{http://www.astro.yale.edu/demarque/yyiso.html}}. The choice of using these particular isochrones was based on the fact that at the time of analysis they were available  with enhanced [$\alpha$/Fe] abundances while the MIST isochrones\footnote{\url{https://waps.cfa.harvard.edu/MIST/}} were only available with solar abundances. Later also MIST has published isochrones with enhanced [$\alpha$/Fe] . 

Figure\,4 in \citet{2025NatAs...9..101X}, which can be found in their \textit{Section on Extended Data}, shows that the ages are very similar, with the MIST ages being slightly higher than those derived with Yonseii-Yale isochrones. Both are equally precise. Using wide binaries \citet{2025arXiv251008675S} evaluated the ages from \citet{2025NatAs...9..101X}  and found that they have high precision. Finally,  \citet{2024A&A...692A.243C} derived stellar ages for a sample of sub-giant and main-sequence turn-off stars from  LAMOST\,DR8 and compared stars in overlap with  \citet{2025NatAs...9..101X}. They used the Stellar Parameters INferred Systematically (SPInS\footnote{\url{https://gitlab.obspm.fr/dreese/spins}}) code to derive ages using the aSTI\footnote{\url{http://basti-iac.oa-abruzzo.inaf.it}} isochrone grid. $\alpha$-enhancement was invoked using the formula from \citet{1993ApJ...414..580S}.  \citet{2024A&A...692A.243C}  found that overall the agreement is good but for some young stars (2-4 Gyr) there is a discrepancy. We are not concerned with stars as young as this in our current study.

In the following sections in this Appendix we discuss the quality cuts we make (Sect.\,\ref{app:cuts}), how we remove the {\it Gaia}-Enceladus from the full sample (Sect.\,\ref{app:gse}), and the influence of imposing a cut at $z=1.5$\,kpc (Sect.\,\ref{app:zcut}).

\subsection{Quality cuts used to select stars for the investigation}
\label{app:cuts}

\begin{figure*}
    \includegraphics[width=16cm]{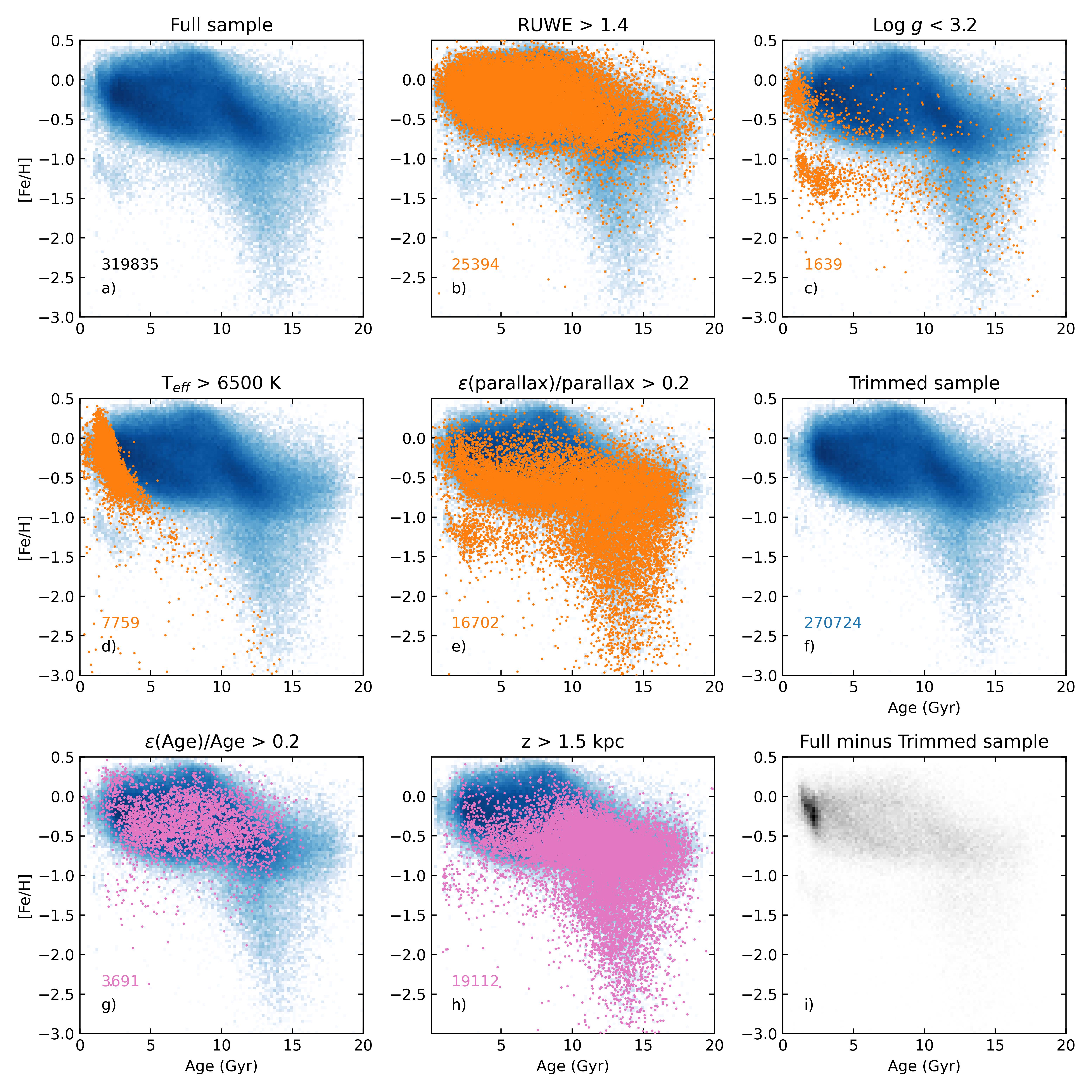}
   \caption{Age-metallicity plots exploring the  properties of the stars in \citet{2025NatAs...9..101X}. {\bf a)} Full sample.  {\bf b)} Shows where the stars with {\tt ruwe} $> 1.4$ fall in orange. {\bf c)} Shows  the stars with $\log g < 3.2$ in orange. {\bf d)}  Shows  the stars with relative error in parallax $>0.2$ in orange. {\bf e)} Shows the sample when it has been trimmed using {\tt ruwe}, parallax and $\log g$. {\bf f)} Shows the trimmed sample with  stars with a relative error in age $>0.2$ in pink. {\bf g)} 2D histogram of the full sample (same as in panel a). In this panel the 2D histogram the maximum value in the grey scale is set to 423. {\bf h)} 2D histogram of the trimmed sample (same as in panel e). In this panel the 2D histogram the maximum value in the grey scale is set to 379. {\bf i)} 2D histogram of the full minus the trimmed sample. In this panel the 2D histogram the maximum value in the grey scale is set to 58. }
\label{fig:quality_cuts_amr}
\end{figure*}

We have investigated the quality of the stellar parameters available in the catalog. In this investigation we came across a group of stars with low metallicities and young ages. Our investigation reveals that these stars  likely have erroneous  ages. They are sitting at the bottom of the red giant branch where it is difficult to determine ages from isochrones \citet{2010ARA&A..48..581S}, \citet{2025arXiv251008675S}. In addition we remove stars with $T_{\rm eff} > 6500$\,K to ensure all ages are well determined.

Figure\,\ref{fig:quality_cuts_amr} explores where stars with certain measurement errors fall in the age-metallicity plane. Panels b) to e) explore the four parameters used to trim the sample ({\tt RUWE}, $\log g$, $T_{\rm eff}$, and parallax) while panel f) shows the trimmed sample. In addition, we show where stars with relative errors in ages of more than 20\% fall (shown in pink in Fig.\ref{fig:quality_cuts_amr} f), 8857 stars) as well as where the stars with $z>1.5$\,kpc fall. Figure\,\ref{fig:quality_cuts_amr} shows the full data-set without any cuts in $z$. In our analysis we apply a cut in $z$ in order to not be influenced by off-plane stars, i.e. halo stars (see Sect.\,\ref{app:zcut}). Based on these assessments, we are reassured that the age determinations from \citet{2025NatAs...9..101X} are overall trustworthy.
The original full sample has 319\,835 stars. Trimming in {\tt RUWE}, $\log g$, $T_{\rm eff}$, and parallax removes 49\,111 stars in total. Figure\,\ref{fig:quality_cuts_amr}  i) shows the difference between the full sample (panel a) and the trimmed sample (panel f).

\subsection{Removing the {\it Gaia}-Enceladus stars from the full sample}
\label{app:gse}

We remove  stars belonging to {\it Gaia}-Enceladus in order to avoid confusion between the existing Milky Way stars  and a major accreted stellar population with distinct kinematic properties. \citet{2022MNRAS.514..689B} in their investigation of the spin-up of the Milky Way disk used elemental abundances (especially [Al/Fe], from APOGEE DR\,17) to remove the accreted component made up by the {\it Gaia}-Enceladus. We do not have access to the style of elemental abundances \citet{2022MNRAS.514..689B} had, instead we remove the stars with the purest definition of the {\it Gaia}-Enceladus in the kinematic space spanned by $L_z$ and $\sqrt{J_R}$  \citep[$|L_z| <500$  kpc\,km\,s$^{-1}$ and $\sqrt{J_R}>30$ (kpc\,km\,s$^{-1}$)$^{1/2}$][]{2021MNRAS.508.1489F,2024MNRAS.527.2165C}. After this removal we inspected the resulting [$\alpha$/Fe] vs [Fe/H] trend. If the {\it Gaia}-Enceladus stars were poorly removed, the well-characterised trend of this population would be visible in this plot as they populate a different trajectory than the Milky Way. We found that the {\it Gaia}-Enceladus was well removed. There is inevitably some contamination from the halo and/or thick disk, but judging from this figure, the contamination is minimal. We have looked to see how effectively the {\it Gaia}-Enceladus stars  were removed by studying the velocity dispersions as a function of galacto-centric radius. The data-set is not really big enough to draw any conclusions about radial differences but we note that the prescription we used to remove the {\it Gaia}-Enceladus stars leaves a small number of such stars in the outer disk. However, their influence on the total data-set is negligible. 

\subsection{Check on the cut in $z$}
\label{app:zcut}

\begin{figure*}
    \includegraphics[width=17cm]{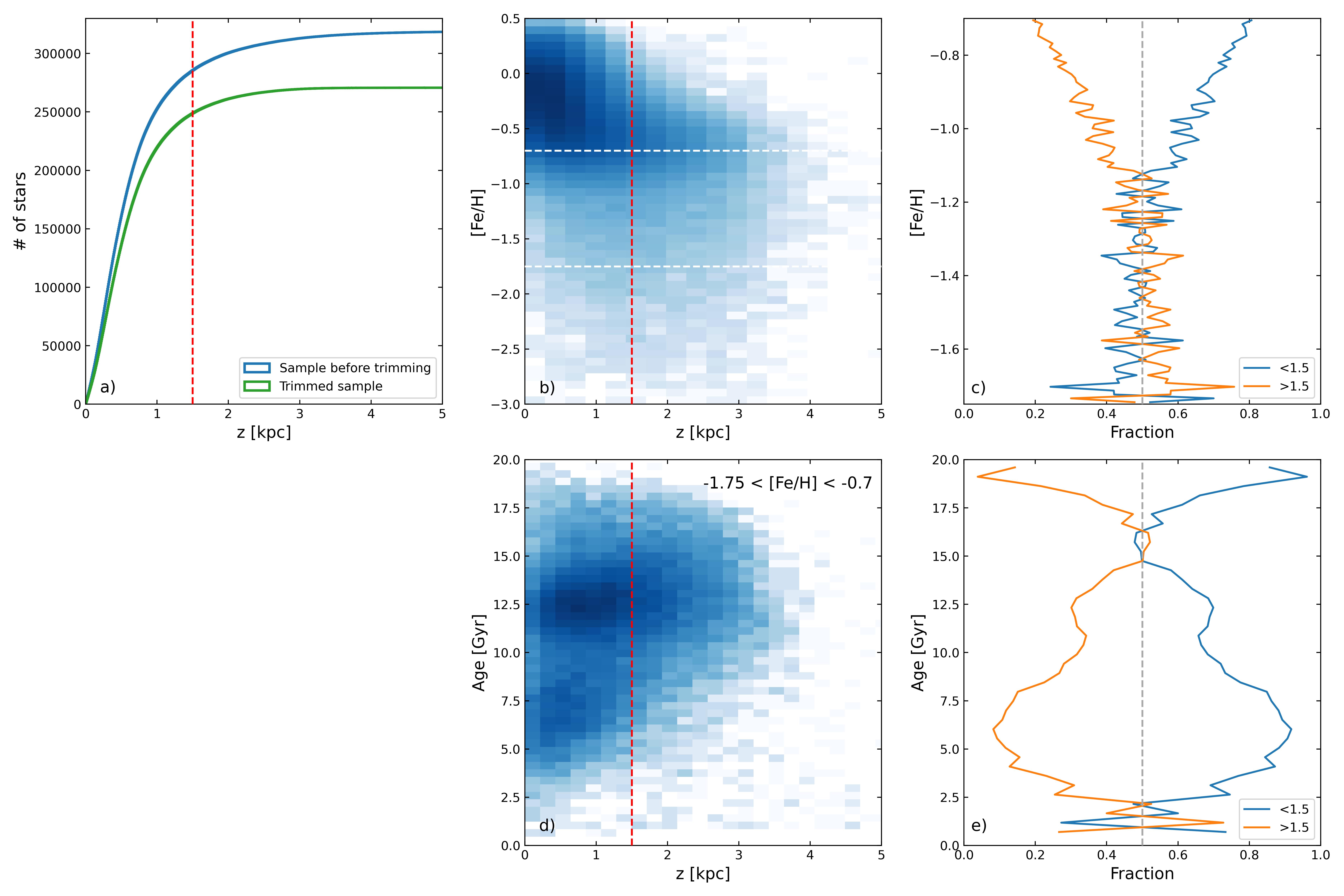}
   \caption{{\bf a)} Cumulative histograms of the stars in the full sample and the trimmed sample (see Sect.\,\ref{sect:data} and App.\,\ref{app:cuts}) as a function of hight above the Galactic plane ($z$). The cut at $z=1.5$\,kpc is marked with a red dashed line. {\bf b)} 2D histogram of [Fe/H] as a function of $z$ for the trimmed sample, where the colour shows the number of stars. The cut at $z=1.5$\,kpc is marked with a red dashed line. The white, horizontal dashed lines mark the [Fe/H]-range over which the Milky Way shows the spin-up, compare Figs.\,\ref{fig:rotsup} and \ref{fig:spinup}. {\bf c)} The fraction of stars above and below $z=1.5$\,kpc  as a function [Fe/H] for the stars with $-1.75 <$[Fe/H]$<-0.7$.  {\bf d)}  2D histogram of Age as a function of $z$ for the stars with $-1.75 <$[Fe/H]$<-0.7$. The colour shows the number of stars. The cut at $z=1.5$\,kpc is marked with a red dashed line. {\bf e)} The fraction of stars above and below $z=1.5$\,kpc  as a function $Age$ for the stars with $-1.75 <$[Fe/H]$<-0.7$. }
\label{fig:zcut}
\end{figure*}

Figure\,\ref{fig:zcut} explores the consequences of the cut we apply at $z=1.5$\,kpc. Figure\,\ref{fig:zcut}a shows the cumulative distribution of the un-trimmed and the trimmed sample (see Sect.\,\ref{sect:data} and App.\,\ref{app:cuts}). It shows that the total number of objects cut out by the cut in $z$ is small in relation to the full sample. 

Figures\,\ref{fig:zcut}b and c explores what happens to the distribution in [Fe/H] when we apply this cut. We find that the cut has little, if any, impact on the [Fe/H] distribution around the [Fe/H] where the Milky Way spins-up. In the Fig.\,\ref{fig:zcut} c we show the fraction of stars above and below the cut (relative to the total number of stars) as a function of [Fe/H] for stars with $-1.75 <$[Fe/H]$<-0.7$.  We see that in the crucial region where the spin-up occurs the stars below the cut in $z$ dominates. 

Next we look at the  distribution of $Age$ and $z$ for only the stars with $-1.75 <$[Fe/H]$<-0.7$. Figure\,\ref{fig:zcut}d shows the 2D histogram. We see two distinct regions, one centred around 12.5\,Gyr and a younger one around 6\,Gyr. It is clear that the older region is dominated by the stars below $z=1.5$\,kpc. This is confirmed when we look at the fraction in  Fig.\,\ref{fig:zcut}e.

From these figures we conclude that we are removing stars in the crucial regions of [Fe/H] and $Age$ but the main fraction of stars are still present in our data set, in particular for the Age determination.

\section{Different ways to calculate the trends and 2D histograms}

\begin{figure*}
\includegraphics[width=18cm]{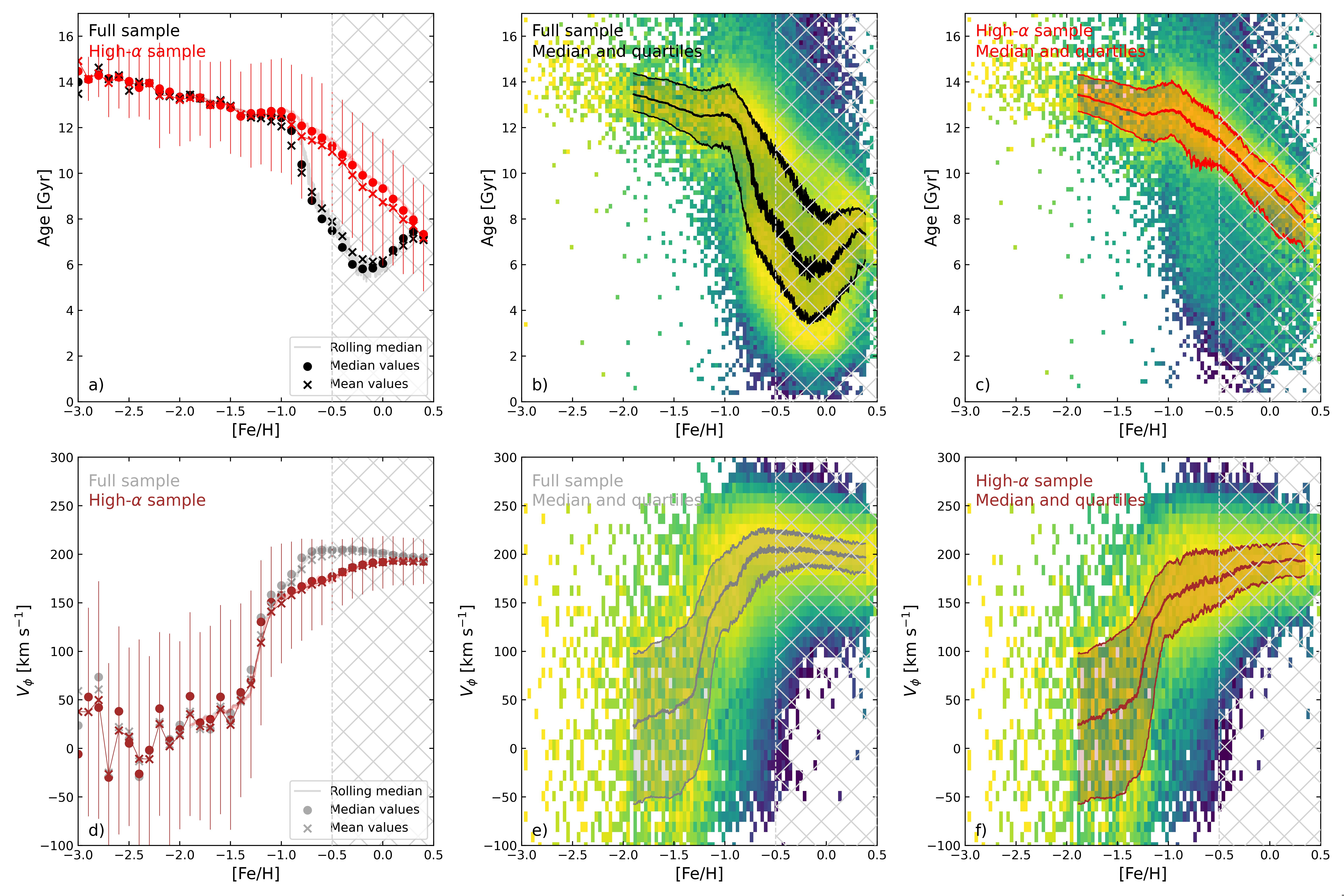}
\caption{
{\bf a)} $Age$ -- [Fe/H] relation. Comparison of the rolling median trend using 1000 datapoints and used in this paper with median and mean values calculated in bins of size 0.1\,dex. Symbols as given in the legend. The errorbars represent the one $\sigma$ around the mean values. For clarity we only show the errorbars for the high-$\alpha$ sample. {\bf b)} $Age$ -- [Fe/H] relation for the full sample as well as a 2D histogram of the whole data-set. We also show the upper and lower quartiles. {\bf c)} $Age$ -- [Fe/H] relation for the high-$\alpha$ sample as well a 2D histogram of the data. We also show the upper and lower quartiles.{\bf d)} $V_{\phi}$ -- [Fe/H] relation. Comparison of the rolling median trend using 1000 datapoints and used in this paper with median and mean values calculated in bins of size 0.1\,dex. Symbols as given in the legend. The errorbars represent the one $\sigma$ around the mean values. For clarity we only show the errorbars for the high-$\alpha$ sample. {\bf e)} $V_{\phi}$ -- [Fe/H] relation for the full sample as well as a 2D histogram of the whole data-set. We also show the upper and lower quartiles. {\bf f)} $V_{\phi}$ -- [Fe/H] relation for the high-$\alpha$ sample as well a 2D histogram of the data. We also show the upper and lower quartiles. In all panels the hatched area shows the data we do not take into account in our analysis as the separation of the high- and low-$\alpha$ stars is less certain above --0.5\,dex. }
\label{fig:spinup2D}
\end{figure*}

In our investigation we have used running medians with a window of 1000 datapoints along the [Fe/H]-axis to calculate the $Age$ -- [Fe/H] and $V_{\phi}$ -- [Fe/H] relations. This gives a robust determination of the different trends. In this Section we inspect it in more detail. 

First we compare the running medians with  medians and means calculated by binning the data by 0.1\,dex. This is shown in Fig.\,\ref{fig:spinup2D} a) and d). The three trends differ slightly but overall they show the same trends. The most notable difference is that for the high-$\alpha$ sample the median $Age$ -- [Fe/H] trend is about 0.5 Gyr higher than the mean trend. We note, however that the stalling of the trend does not change. The difference is perhaps more pronounced around the actual spin-up in the$V_{\phi}$ -- [Fe/H]  diagram. However, that spin-up remains aligned with the stalling of the age trend. 

In Fig. Fig.\,\ref{fig:spinup2D} b), c), e) and f) we show the corresponding 2D histograms (column normalized) for the $Age$ -- [Fe/H] and $V_{\phi}$ -- [Fe/H] relations. In each we also  the running median and its associated quartiles. From this comparison we can conclude that the running median and associated quartiles do a good  reproducing the underlying 2D histograms.

In all plots we have hatched the areas for [Fe/H] $>$ --0.5\,dex to emphasise that we do not use these data to draw any conclusions. The split between high- and low-$\alpha$ stars is difficult above --0.5\,dex. 


\bibliography{Feltzing_AAS75220}{}
\bibliographystyle{aasjournal}



\end{document}